\documentclass[aip, pop, preprint]{revtex4-1}
\usepackage{graphicx}
\usepackage{color}
\usepackage{amssymb}
\usepackage{amsmath}
\usepackage{tabularx}
\usepackage{bm}
\usepackage{hyperref}
\usepackage{ltxtable}
\usepackage{natbib} 
\usepackage{longtable}
\usepackage{natbib}
\usepackage[usenames,dvipsnames]{xcolor}
\hypersetup{
colorlinks,
citecolor=blue,
filecolor=blue,
linkcolor=blue,
urlcolor=blue}

\newcommand{\be}{\begin{equation}}
\newcommand{\ee}{\end{equation}} 
\newcommand{\bea}{\begin{eqnarray}}
\newcommand{\eea}{\end{eqnarray}}
\newcommand{\la}{\langle}
\newcommand{\ra}{\rangle}

\begin{document}

\title{Taylor's Frozen-in Hypothesis for Magnetohydrodynamic turbulence and Solar Wind}
%\title{Similarities between the structure functions of thermal convection and hydrodynamic turbulence}

\author{Mahendra K. Verma}
\email{mkv@iitk.ac.in}
\affiliation{Department of Physics, Indian Institute of Technology Kanpur, Kanpur 208016, India}

\date{1 March 2022}%
%\revised{August 2010}%

\begin{abstract}
In hydrodynamics, Taylor's frozen-in hypothesis connects the wavenumber spectrum to the frequency spectrum of a time series measured in real space.  In this paper, we generalize Taylor's hypothesis to magnetohydrodynamic turbulence. We analytically derive one-point two-time correlation functions for Els\"{a}sser variables whose Fourier transform yields the corresponding frequency spectra, $ E^\pm(f) $. We show that $ E^\pm(f) \propto |{\bf U}_0 \mp {\bf B}_0|^{2/3}  $   in Kolmogorov-like model, and $ E^\pm(f) \propto (B_0  |{\bf U}_0 \mp {\bf B}_0|)^{1/2}  $ in Iroshnikov-Kraichnan model, where $ {\bf U}_0, {\bf B}_0$ are the mean velocity and mean magnetic fields respectively.  
\end{abstract}
%\pacs{47.27.te, 47.27.-i, 47.55.P-}
\maketitle

\section{Introduction}

Theoretical, experimental, and computational tools are employed for studying turbulence.    \citet{Kolmogorov:DANS1941Dissipation,Kolmogorov:DANS1941Structure} constructed one of the most popular model of turbulence and showed that for homogeneous and isotropic turbulence, the kinetic energy spectrum in wavenumber space is 
\be
E_u(k) = K_\mathrm{Ko} \epsilon_u^{2/3} k^{-5/3},
\label{eq:Kolm_spectrum}
\ee
where $ k $ is the wavenumber, $ \epsilon_u $ is the  dissipation rate of kinetic energy, and $ K_\mathrm{Ko}  $ is  Kolmogorov's constant. Computation of  $ E_u(k) $ requires three-dimensional velocity field, which is difficult to measure in experiment. Instead, the velocity field is measured at select real-space points.   \citet{Taylor:PRS1938}  proposed an important conjecture that helps connect $ E_u(k)  $  to the frequency spectrum, $ E_u(f) $,  of the measured time series. \citet{Taylor:PRS1938} hypothesized that turbulent fluctuations are advected by the mean flow (velocity $ {\bf U}_0 $) as if fluctuations are frozen in the flow. Under this assumption, using Kolmogorov's spectrum of Eq.~(\ref{eq:Kolm_spectrum}), we can easily derive the  frequency spectrum as
\be
E_u(f) = A (\epsilon_u U_0)^{2/3} f^{-5/3},
\label{eq:Taylor_Ef}
\ee
where $ A $ is a constant. Thus, Taylor's hypothesis plays an important role in turbulence experiments and data analysis.

\citet{Kraichnan:PF1964Eulerian} argued that the large-scale eddies too sweep the small-scale fluctuations, a phenomenon  called \textit{sweeping effect}. \citet{Wilczek:PRE2012} and  \citet{He:ARFM2017}  generalized  Eq.~(\ref{eq:Taylor_Ef}) by  taking into account  the sweeping effect.   Recently, \citet{Verma:INAE2020_sweeping} performed a detailed calculation of  temporal correlation function that includes sweeping effect and turbulent diffusion; they showed that $ E_u(f) \propto f^{-5/3} $ in the presence of a mean flow, but $ E_u(f) \propto f^{-2} $ when $ U_0 $ is small compared to the fluctuations.

In magnetohydrodynamic (MHD) turbulence, the velocity and magnetic fields have their mean values ($ {\bf U}_0 $ and ${\bf B}_0 $, respectively, in velocity units) and fluctuations. Hence, Taylor's hypothesis and Eq.~(\ref{eq:Taylor_Ef}) need  generalization, especially because  Solar wind and solar corona are  important MHD turbulence laboratories where spacecrafts make in situ measurements of the velocity and magnetic fields.   Taylor's hypothesis would help us derive the wavenumber spectrum using the frequency spectrum of the time series.  Note that the typical speed of  spacecrafts are much smaller than the solar wind speed (250 km/s to 750 km/s). Hence, the spacecraft can be assumed to be stationary, consistent with the assumptions of Taylor's hypothesis. Another important factor is that $ B_0 \ll U_0$  at distances 0.3 AU or higher, where most spacecrafts are located. As a result, we can ignore the effects of the mean magnetic field and employ Taylor's hypothesis. In one of the first applications of Taylor's hypothesis to the solar wind, \citet{Matthaeus:JGR1982rugged} deduced $ k^{-5/3} $ spectrum for the solar wind by employing Eq.~(\ref{eq:Taylor_Ef}). There are many more works on solar wind observations, e.g.,~\citet{Podesta:ApJ2007}.

Recent spacecraft \textit{Parker Solar Probe} (PSP) has come quite close the Sun (as close as 8.86 solar radii).   At this distance, $ B_0 $ is comparable to $ U_0 $, and hence, we need to examine whether Taylor's hypothesis can be employed to PSP data.  In addition, the solar wind is anisotropic that  makes an application of Taylor's hypothesis problematic. \citet{Bourouaine:ApJ2018} modelled the Fourier-transformed  temporal correlation function of MHD turbulence using a Gaussian model rather than a pure exponential model, and argued that the decorrelation frequency is  linearly related to the perpendicular wavenumber.   In a related work, \citet{Perez:AA2021} investigated the validity of Taylor's hypothesis for the PSP data, and showed that the  frequency spectrum accurately represents the spectral indices associated with the underlying spatial spectrum of turbulent fluctuations in the plasma frame.  In a another recent work, \citet{Kasper:PRL2021} computed the frequency spectrum of the PSP data recorded at 13 million km above the photosphere, and observed the spectral index  to be closer to $ -3/2 $ than $ -5/3 $. Note that \citet{Kasper:PRL2021} assume Taylor's hypothesis for the solar wind even though $ B_0 $ and $ U_0 $  are comparable in this regime.

In this paper, we extend the derivation of \citet{Verma:INAE2020_sweeping} for hydrodynamic turbulence to MHD turbulence.  We express the ``dressed" or ``renormalized" temporal correlation functions for the Els\"{a}sser variables in the Fourier space. Then, an inverse Fourier transform of the above  correlation function yields two-point two-time correlation function.  By setting the distance between the two points to zero, we deduce one-point two-time correlation functions, whose Fourier transform yields the frequency spectra for MHD turbulence. For  Kolmogorov's and Iroshnikov-Kraichnan models~\cite{Iroshnikov:SA1964,Kraichnan:PF1965MHD}, we obtain respectively $ f^{-5/3} $ and $ f^{-3/2} $ frequency spectra with prefactors that are functions of $| {\bf U}_0  \mp {\bf B}_0| $, which are the wave speeds of Alfv\'{e}n waves.  Thus, we generalize Taylor's hypothesis to MHD turbulence; our derivation is mathematically more rigorous than  earlier ones.  

In the next two sections, we derive the correlation functions for linear and turbulent MHD.

\section{Correlation functions for linear MHD}

The equations of a magnetofluid moving with a mean velocity of ${\mathbf U}_0$ in a background of  a mean magnetic field $ {\bf B}_0 $ are
\begin{eqnarray}
	\frac{ \partial{\bf u}}{\partial t} + ({\mathbf U}_0 \cdot \nabla) {\bf u} - ({\mathbf B}_0 \cdot \nabla) {\bf b}  + ({\bf u} \cdot \nabla) {\bf u} -  ({\bf b} \cdot \nabla) {\bf b} & = &  -\nabla p + \nu \nabla^2  {\bf u} + {\bf f}, 
	\label{eq:u} \\
		\frac{ \partial{\bf b}}{\partial t} + ({\mathbf U}_0 \cdot \nabla) {\bf b} - ({\mathbf B}_0 \cdot \nabla) {\bf u}  + ({\bf u} \cdot \nabla) {\bf b} -  ({\bf b} \cdot \nabla) {\bf u} & = &  \eta \nabla^2  {\bf b}, 
	\label{eq:b} \\
	\nabla \cdot  {\bf u}  = 0,
	~~~\nabla \cdot  {\bf b}  = 0, 
	 \label{eq:divfree}
\end{eqnarray}
where ${\bf u,b}$ are the velocity and magnetic field fluctuations, ${\bf f}$ is the external force,  $p$ is the pressure,  $\nu$ is the kinematic viscosity, and $ \eta $ is the magnetic diffusivity.  In the above equations, the magnetic field is velocity units, which is obtained by a transformation, $ {\bf B}_\mathrm{CGS} \rightarrow {\bf B}_\mathrm{CGS}/\sqrt{4\pi\rho}$, where $ \rho $ is the material density of the flow. In this paper, we assume that the flow is incompressible.  

Alfv\'{e}n waves are the basic modes of linearized MHD equations, and they
 are conveniently expressed in terms of Els\"{a}sser variables, $ {\bf z^\pm = u \pm b} $.  Using Eqs.~(\ref{eq:u}, \ref{eq:b}, \ref{eq:divfree}), we can derive the following equations for  $ {\bf z^\pm}$: 
\bea
	\frac{ \partial{\bf z}^\pm}{\partial t} + {\mathbf Z}_0^\mp \cdot \nabla {\bf z}^\pm    + ({\bf z}^\mp \cdot \nabla) {\bf z}^\pm  & = &  -\nabla p + \nu_\pm \nabla^2  {\bf z}^\pm + \nu_\mp \nabla^2  {\bf z}^\mp  + {\bf f}, 
\label{eq:z} \\
	\nabla \cdot  {\bf z}^\pm  = 0,
\eea
where ${\mathbf Z}_0^\mp =  ({\mathbf U}_0 \mp {\mathbf B}_0)  $, and  $ \nu_\pm = (\nu \pm \eta)/2 $. For simplification, in this paper, we take $ \nu = \eta$ that leads to $\nu_+ = \nu = \eta$ and $\nu_- = 0$.  In Fourier space, the equations for $ {\bf z}^\pm $ are
\bea
\left[\frac{ \partial}{\partial t} + i {\mathbf Z}_0^\mp  \cdot {\bf k} + \nu k^2 \right] {\bf z}^\pm ({\bf k})  
 & = &  -i {\bf k} p({\bf k})   - i \sum_{\bf p}  [{\bf k \cdot z^{\mp} ({\bf q}) ]  z^\pm ({\bf p}) } + {\bf f} ({\bf k}),
\label{eq:zk} \\ 
{\bf k} \cdot  {\bf z}^\pm ({\bf k})  = 0,
\eea
with $ {\bf q = k-p} $.  A linearized version of the above equations indicates that the waves $ {\bf z}^+ ({\bf k})  $ and  $ {\bf z}^- ({\bf k})  $ move with speeds $ ({\mathbf U}_0 - {\mathbf B}_0) \cdot {\bf k}  $ and $({\mathbf U}_0 + {\mathbf B}_0) \cdot {\bf k} $ respectively. 

Using the  linearised versions of Eq.~(\ref{eq:zk}),  we derive the  equations for the Green's functions 
\bea
\left[\frac{ \partial}{\partial t} + i {\mathbf Z}_0^\mp   \cdot {\bf k} + \nu k^2 \right] G^\pm ({\bf k},t,t')    & = & \delta(t-t'),
\label{eq:Green_linear}
\eea
whose solutions are
\be
G^\pm({\bf k},\tau) =\theta(\tau) \exp{[-i  {\mathbf Z}_0^\mp   \cdot {\bf k} \tau]}  \exp{(- \nu k^2 \tau)},
\label{eq:Gkt_nu0}
\ee
where $\tau = t-t'$, and $\theta(\tau)$ is the step function.  

The equal-time correction functions, $C^\pm({\bf k},0)$, and two-time correction functions, $C^\pm({\bf k},\tau)$, for  $ {\bf z^\pm (k)} $ are defined as
\bea
C^\pm({\bf k},0) & = & \langle |\mathbf z^\pm(\mathbf k, t)|^2 \rangle, \\
C^\pm({\bf k},\tau) & = & \langle \mathbf z^\pm(\mathbf k, t)\cdot \mathbf z^{\pm*}(\mathbf k, t+\tau) \rangle.
\label{eq:C_k_tau_def}
\eea
Note that $C^\pm({\bf k},\tau)  $ is a complex function.  We define the  normalised  correlation function as
\begin{equation}
	R^\pm(\mathbf k, \tau) = \frac{C^\pm(\mathbf k, \tau)}{C^\pm(\mathbf k, 0)}.
	\label{eq:R}
\end{equation} 
A   generalisation of fluctuation-dissipation theorem to MHD yields~\cite{Kiyani:PRE2004}
\be
R^\pm(\mathbf k, \tau) = G^\pm({\bf k},\tau) =\theta(\tau) \exp{(-i {\mathbf Z}_0^\mp  \cdot { \bf k}   \tau)}  \exp{(- \nu k^2 \tau)}.
\label{eq:Rkt_linear}
\ee
The above equation indicates that the normalised correlation functions exhibit damped oscillations.

\section{Correlation functions for turbulent MHD}

A magnetofluid becomes turbulent when $ UL/\nu \gg 1  $ and $ UL/\eta \gg 1 $, where $ U, L $ are the large-scale velocity and length respectively.  There is no definitive theory of MHD turbulence, rather it has many models~\cite{Verma:PR2004,Verma:book:ET,Beresnyak:LR2019}. In this paper we will focus on two leading models. Here, we focus on the shell spectra: $ E^\pm(k) $, $ E_u(k) $, $ E_b(k) $,  which are defined as
\bea
E^\pm   & = & \frac{1}{2} \la |{\bf z^\pm|^2} \ra = \int  E^\pm(k) dk, \\
E_u   & = & \frac{1}{2} \la |{\bf u}^2 \ra = \int  E_u(k) dk, \\
E_b   & = & \frac{1}{2} \la |{\bf b}^2 \ra = \int  E_u(k) dk,
\eea
where $ E^\pm, E_u , E_b  $ are the total energies per unit volume of $ {\bf z}^\pm $, $ {\bf u} $, and $ {\bf b} $  respectively.

\begin{enumerate}
	\item {\em Komogorov-like MHD turbulence phenomenology}: In this framework, the energy spectra $ E^\pm(k) $ are modelled as~\cite{Marsch:RMA1991,Verma:JGR1996DNS,Verma:PR2004}
	\be
	E^\pm(k) = K^\pm (\epsilon^\pm)^{4/3} (\epsilon^\mp)^{-2/3} k^{-5/3},
	\ee
	where $ \epsilon^\pm $ are the inertial-range energy fluxes or dissipation rates of $ {\bf z}^\pm $, and $ K^\pm $ are constants, similar to Kolmogorov's constant for hydrodynamic turbulence. This phenomenology is also referred to as \textit{imbalanced MHD}. 
	
	In addition, \citet{Goldreich:ApJ1995} constructed a phenomenology for the anisotropic MHD turbulence.  Using \textit{critical balance} between the  time scales for the nonlinear interactions and Alfv\'{e}n wave propagation, they showed that the modal energy
	\be
	\tilde{E}(k_\perp,k_\parallel) = K \epsilon^{2/3} k_\perp^{-10/3} g(k_\parallel/k_\perp^{2/3}),
	\label{eq:Ek_GS}
\ee
	where $ K $ is a constant, $ \epsilon $ is the total dissipation rate, and $k_\parallel$ and $k_\perp$ are respectively the wavenumber components parallel and perpendicular to the mean magnetic field. Note that
	\be
	\int k_\perp dk_\perp dk_\parallel 	\tilde{E}(k_\perp,k_\parallel)  = E,
	\ee
	where $ E$ is the total energy.
		                                                                 
	\item {\em Iroshnikov-Kraichnan phenomenology\cite{Iroshnikov:SA1964,Kraichnan:PF1965MHD}}: In this framework, the Alfv\'{e}n time scale, $ (k B_0)^{-1} $, is the relevant time scale, leading to the energy spectrum as 
	\be
	E_u(k) \approx E_b(k) \approx K_\mathrm{IK} (\epsilon B_0)^{1/2} k^{-3/2},
	\ee	
    where $ K_\mathrm{IK} $ is constant, and $ B_0 $ is the amplitude of the mean magnetic field or that of large-scale magnetic field. In this phenomenology, the	kinetic and magnetic energies are equipartitioned.  
	
	\citet{Dobrowolny:PRL1980} showed that 
	\be
	\epsilon^+ = \epsilon^- = \frac{1}{B_0} E^+(k) E^-(k) k^3.
	\ee
	For a special case when $ E^+(k) = E^-(k) =E(k) $ ($ E$ is the total energy), we obtain
	 \be
	 E^+(k) = E^-(k)  = E(k) = K' (\epsilon B_0)^{1/2}  k^{-3/2}.
	 \label{eq:E(k)_MHD_IK}
	 \ee	   
\end{enumerate}

Now, we model the correlation function for MHD turbulence following the strategies adopted for  the hydrodynamic turbulence. The most critical part is the  convective component. Using Eq.~(\ref{eq:zk}), we deduce the convective component to be  
$ \exp{(-i {\mathbf Z}_0^\mp  \cdot { \bf k}   \tau)} $ for $ {\bf z}^\pm $. In the following discussion we show that the convective part contributes most significantly to the frequency spectra. 

The other two parts of the correlation function are the effective diffusion parameters and the sweeping effect. 
In hydrodynamic turbulence, field-theoretic treatment shows that the \textit{renormalized viscosity}  or \textit{effective viscosity} ($ \nu(k) $) is
\be
\nu(k) = \nu_* \epsilon^{1/3} k^{-4/3},
\label{eq:nu(k)_HD}
\ee
where $ \nu_* $ is a nondimensional constant\cite{Yakhot:JSC1986,	McComb:book:Turbulence,Verma:PR2004}. However, MHD turbulence has two diffusion parameters, viscosity and magnetic diffusivity, that depend on the cross helicity, Alfv\'{e}n ratio, and mean magnetic field. We do not yet have general formulas for these renormalized parameters, even though they have been solved for special cases~(see \citet{Verma:PRE2001,Verma:Pramana2003Nonhelical,Verma:Pramana2003Helical}). In this paper, we  simplify the calculation by assuming that both the renormalized parameters are equal (i.e., $ \nu(k) = \eta(k)  $), and that for Kolmogorov-like phenomenology,
\be
\nu^\pm(k) = \nu^\pm_* (\epsilon^\pm)^{1/3} k^{-4/3},
\label{eq:nu(k)_MHD}
\ee
and for Iroshnikov-Kraichnan phenomenology,
\be
\nu^\pm(k) = \nu'_* (\epsilon B_0)^{1/4} k^{-5/4}.
\label{eq:nu(k)_IK}
\ee
Here, $\nu^\pm_*  $ and $ \nu'_* $ are constants.  As we show in the next section, the terms with $ \nu^\pm(k) $ get integrated in $ E^\pm(f) $. Hence, a precise form of  $ \nu^\pm(k)  $ may not be critical for the derivation of $ E^\pm(f) $. 

In addition, according to the \textit{sweeping effect}, large-scale flow structures sweep the inertial-range fluctuations. For hydrodynamic turbulence, \citet{Kraichnan:PF1964Eulerian}, \citet{Wilczek:PRE2012}, \citet{Verma:INAE2020_sweeping} and others have constructed models for the sweeping effect.  For MHD turbulence, we follow the prescription of \citet{Verma:INAE2020_sweeping} who added a random large-scale velocity field, $ \tilde{\bf U}_0 $, to the mean velocity field $ {\bf U}_0 $. These corrections are added in the correlation function for the linear equation [Eq.~(\ref{eq:Rkt_linear})].

Under the above assumptions, we arrive at the following expressions for the correlation functions of MHD turbulence~\cite{Verma:INAE2020_sweeping}: 
\bea
R^\pm (\mathbf k, \tau) = \frac{C^\pm (\mathbf k, \tau)}{C^\pm (\mathbf k)} & = &  \exp{(-i {\mathbf Z}_0^\mp  \cdot { \bf k}   \tau)}\exp(-i  {\bf \tilde{U}_0 \cdot k}  \tau) \exp[-\nu^\pm(k) k^2 \tau] . 
\label{eq:Rk_MHD}
\eea
Note that the correlations functions depend on both  ${\bf U}_0$ and ${\bf B}_0$. In the next section, we will relate the above functions to Taylor's hypothesis for MHD turbulence.

%%%%%
\section{Taylor's hypothesis for MHD turbulence}
\label{sec:Taylor_hypo}
Using  Eq.~(\ref{eq:Rk_MHD}), we first derive the two-point two-time correlation functions, after that we derive one-point two-time correlation functions, whose Fourier transform yields the frequency spectra of real-space time series of $ {\bf z}^\pm $.

Using  Eq.~(\ref{eq:Rk_MHD}),  we derive the following two-point two-time correlation functions for $ {\bf z}^\pm $:
\bea
C^\pm({\bf r}, \tau) & = & \int d{\bf k} C^\pm({\bf k}) \exp[-\nu(k) k^2 \tau-i{\bf Z^\mp_0 \cdot k} \tau] \exp[-i {\bf k} \cdot \tilde{\bf U}_0({\bf k}) \tau] \exp(i {\bf k} \cdot {\bf r}),
\label{eq:corr_U0}
\eea 
where $  {\bf Z}^\mp_0 =  {\bf U}_0 \mp  {\bf B}_0$.  We ensemble average $C^\pm({\bf r}, \tau)$ for  random  $ \tilde{\bf U}_0 $ (assuming isotropic, as in \citet{Kraichnan:PF1964Eulerian}) that yields~\cite{Kraichnan:PF1964Eulerian, Wilczek:PRE2012,Verma:INAE2020_sweeping}
\bea
C^\pm({\bf r}, \tau) & = & \int d{\bf k} C^\pm({\bf k}) \exp[-\nu^\pm(k) k^2 \tau - i{\bf Z^\pm_0 \cdot k} \tau] \langle \exp[-i c k \tilde{U}_0( k)\tau]  \rangle \exp(i {\bf k} \cdot {\bf r}) \nonumber \\
& = & \int d{\bf k} C^\pm({\bf k})\exp[-\nu^\pm(k) k^2 \tau - i{\bf Z^\pm_0 \cdot k} \tau]    \exp[- c^2 k^2  \{\tilde{U}_0( k) \}^2 \tau^2] \exp(i {\bf k} \cdot {\bf r}).
\label{eq:Cpm_r_t}
\eea 
For simplicity, we assume that the constant $ c \approx 1 $, and   set ${\bf r}=0$ to compute  one-point two-time correlation functions  $ C^\pm({\bf r}=0, \tau) =  C^\pm(\tau)$. The Gaussian model for the sweeping effect has been reported earlier by \citet{Kraichnan:PF1964Eulerian,Wilczek:PRE2012}, and \citet{Bourouaine:ApJ2018}.

Now we derive $ C^\pm(\tau)  $ for the Kolmogorov-like phenomenology. Following Pope~\cite{Pope:book}, we take  $C^\pm({\bf k})$:
\be
C^\pm({\bf k})  = \frac{E^\pm(k)}{2\pi k^2} =  \frac{1}{2\pi k^2} f_L(kL) f_\eta(k\eta) K^\pm k^{-5/3} \frac{(\epsilon^\pm)^{4/3}}{(\epsilon^\mp)^{2/3}}  
\label{eq:Ek_Pope}
\ee
where  
\begin{eqnarray}
	f_L(kL) & = & \left( \frac{kL}{[(kL)^2 + c_L]^{1/2}} \right)^{5/3+p_0}, 
	\label{eq:fL} \\
	f_\eta(k\eta) & = & \exp \left[ -\beta \left\{ [ (k \eta)^4 + c_\eta^4 ]^{1/4}   - c_\eta \right\} \right]
	\label{eq:feta}
\end{eqnarray}
are respectively the forcing and dissipative  components of the energy spectra, and  $c_L, c_\eta, p_0, \beta$ are constants.  We employ Eq.~(\ref{eq:nu(k)_MHD}) for $ \nu(k) $, and   $\tilde{U}_0(k) = \epsilon^{1/3}k^{-1/3}$ with $ \epsilon $ as the total dissipation rate (see \citet{Verma:INAE2020_sweeping}).  In addition, we ignore   the constants for brevity.   After the above substitutions in Eq.~(\ref{eq:Cpm_r_t}) with $ {\bf r}=0 $, we obtain
\bea
C^\pm(\tau) & = &  K^\pm \frac{(\epsilon^\pm)^{4/3}}{(\epsilon^\mp)^{2/3}}    \int dk  k^{-5/3} f_L(kL) f_\eta(k\eta) \exp(-i{\bf Z^\mp_0 \cdot k} \tau)  \times  \nonumber \\
&&  \exp[-(\epsilon^\pm)^{1/3}k^{2/3}\tau]  \exp[-\epsilon^{2/3}k^{4/3}\tau^2].
\label{eq:C_tau_appendix}
\eea 
The above integral is quite complex, but it can be simplified in the  asymptotic case.
 
 For Taylor's frozen-in hypothesis to work, we  assume that  
 $\mathbf Z^\mp_0 \cdot \mathbf k \gg \nu^\pm(k) k^2$ and $\mathbf Z^\mp_0 \cdot \mathbf k \gg k \tilde{U}_0(k)$.   In addition, for $ C^+(\tau) $ and $ C^-(\tau) $,   we choose the $z$ axis to be along the direction of ${\bf Z}^-_0$ and ${\bf Z}^+_0$ respectively.  For simplification, we  make a change of variable, $\tilde{k}_\pm =   Z^\mp_0 k \tau$, and use $ \epsilon = U^2/T $ ($ U $ is the rms speed).   As a result, we obtain  
\bea
C^\pm(\tau) & \approx &  K^\pm ( Z^\mp_0 \tau)^{2/3} \frac{(\epsilon^\pm)^{4/3}}{(\epsilon^\mp)^{2/3}}    \int d{\tilde{k}_\pm} \tilde{k}_\pm^{-5/3} f_L[\tilde{k}_\pm (L/Z^\mp_0\tau)]  f_\eta[\tilde{k}_\pm (\eta / U_0 \tau)]  \frac{\sin(Z^\mp_0 k \tau)}{Z^\mp_0 k \tau} \times   \nonumber \\
&&  \exp[- \tilde{k}_\pm^{2/3} (U/Z^\mp_0)^{2/3} (\alpha^\pm \tau/T)^{1/3} - \tilde{k}_\pm^{4/3} (U/Z^\mp_0)^{4/3} (\tau/T)^{2/3} ],
\eea 
where $\epsilon^\pm = \epsilon \alpha^\pm $.  For $\tau$ in the inertial range,  $L/Z^\mp_0\tau \gg 1$ and $\eta/ Z^\mp_0 \tau \ll 1$.  Consequently,  $f_L(\tilde{k}_\pm (L/Z^\mp_0\tau)) \approx 1$, and $f_\eta( \tilde{k}_\pm (\eta / Z^\mp_0 \tau) \approx 1$.     Therefore,
\bea
C^\pm(\tau) & \approx & K^\pm  ( Z^\mp_0 \tau)^{2/3} \frac{(\epsilon^\pm)^{4/3}}{(\epsilon^\mp)^{2/3}}   \int d{\tilde{k}_\pm} \tilde{k}_\pm^{-5/3} \frac{\sin \tilde{k}_\pm}{\tilde{k}_\pm } \times \nonumber \\
&&  \exp[- \tilde{k}_\pm^{2/3} (U/Z^\mp_0)^{2/3} (\alpha^\pm \tau/T)^{1/3} - \tilde{k}_\pm^{4/3} (U/Z^\mp_0)^{4/3} (\tau/T)^{2/3} ] \nonumber \\ 
& = & A^\pm  K^\pm ( Z^\mp_0 \tau)^{2/3} \frac{(\epsilon^\pm)^{4/3}}{(\epsilon^\mp)^{2/3}},
\eea 
where $A^\pm$ are the values of the nondimensional integrals.  The Fourier transform of the above $ C^\pm(\tau)$ yields the following frequency spectra:
\bea
E^\pm(f) & \approx & \int C^\pm(\tau) \exp(-i 2\pi f \tau) d\tau = \int  A^\pm K^\pm ( Z^\mp_0 \tau)^{2/3} \frac{(\epsilon^\pm)^{4/3}}{(\epsilon^\mp)^{2/3}} \exp(-i 2\pi f \tau) d\tau  \nonumber \\
& =  & A'^\pm ( |{\bf U}_0 \mp {\bf B}_0|)^{2/3}  \frac{(\epsilon^\pm)^{4/3}}{(\epsilon^\mp)^{2/3}} f^{-5/3},
\label{eq:MHD_Taylor_Kolm}
\eea
where $ A'^\pm $ are constants. Thus, we obtain $ -5/3 $ frequency spectra for Kolmogorov-like MHD turbulence phenomenology.   Note that $E^+(f)$  and $E^-(f)$ are functions of $ |{\bf U}_0 - {\bf B}_0| $ and $ |{\bf U}_0 + {\bf B}_0 |$, respectively, which are the respective speeds of  $ {\bf z^+(k)}$ and $ {\bf z^-(k)}$ in the linear approximation. In comparison to Eq.~(\ref{eq:Taylor_Ef}), $ E^\pm(f) $ has $ U_0 \rightarrow  |{\bf U}_0 \mp {\bf B}_0| $ and 
$ \epsilon_u \rightarrow (\epsilon^\pm)^{4/3} (\epsilon^\mp)^{-2/3}$.

The calculation for the anisotropic MHD turbulence is more complex. However, the complexity is likely to be in the integral computation, which will reflect in the constants $ A^\pm $. For anisotropic MHD turbulence, we expect that the form  of $ E^\pm(f) $ will  the  same as in Eq.~(\ref{eq:MHD_Taylor_Kolm}).

The above analysis can be  extended to Iroshnikov-Kraichnan phenomenology, where the correlation functions   are
\be
C^\pm({\bf k})  = \frac{E^\pm(k)}{2\pi k^2} \sim  
  ( \epsilon B_0)^{1/2} k^{-7/2}.
\label{eq:Ck_Kraichnan}
\ee
We substitute Eq.~(\ref{eq:Ck_Kraichnan}) into Eq.~(\ref{eq:Cpm_r_t}), employ $ \nu^\pm(k) $ of Eq.~(\ref{eq:nu(k)_IK}), and set $ {\bf r} = 0$. Following the same steps as above, we obtain
\bea
C^\pm(\tau)  \approx   (\epsilon B_0 Z^\mp_0 \tau)^{1/2}.
\eea
 Fourier transform of the above $ C^\pm(\tau)  $ yields the following frequency spectra:
\bea
E^\pm(f)  = A_\mathrm{IK}  (\epsilon B_0  |{\bf U}_0 \mp {\bf B}_0| )^{1/2} f^{-3/2},
\label{eq:MHD_Taylor_Kr}
\eea
where $ A_\mathrm{IK} $  is a constant. Thus, we obtain $ f^{-3/2} $ frequency spectrum for the Iroshnikov-Kraichnan phenomenology.  

The prefactors for $ E^\pm(f) $ are functions of $ Z^\pm_0 $. However, the prefactors  for $ E_u(k) $ and $ E_b(k) $ would be more complex because
\be
E^\pm(f) = E_u(f) + E_b(f)  \pm 2 H_c(f),
\ee
where $ H_c = (1/2) \la {\bf u \cdot b} \ra$. Clearly,  derivation of $ E_u(k)$, $ E_b(k)$, and $H_c(k) $ requires further inputs, e.g., relationships among these functions.  Based on these complexities,

\section{Discussion and Conclusions}

In this paper, we extend Taylor's frozen-in hypothesis to MHD turbulence. From the first principle,  we derive one-point two-time correlation functions for MHD turbulence, whose Fourier transform  yields the corresponding frequency spectra.  The main predictions of our quantitative calculations are as follows:
\begin{enumerate}
	\item The spectral indices for $ E^\pm(k) $ and $ E^\pm(f) $ are the same.
	\item The prefactors of $E^\pm(f) $ are proportional to $ |{\bf U_0 \mp B_0}|^{2/3 }$ in  Kolmogorov-like phenomenology, but proportional to $ B_0^{1/2} |  {\bf U_0 \mp B_0}|^{1/2 }$ in Iroshnikov-Kraichnan phenomenology.  In contrast to $ E_u(f) $ for hydrodynamic turbulence, $ U_0  \rightarrow|  {\bf U_0 \mp B_0}| $ in $E^\pm(f) $ of MHD turbulence.
	
	\item The kinetic and magnetic energy spectra, $ E_u(f) $ and $ E_b(f) $,  are expected to have more complex prefactors.  
	
	\item When $ B_0 \ll U_0 $, the frequency spectrum of Eq.~(\ref{eq:Taylor_Ef}) can be employed for all the fields.  
\end{enumerate}

The above predictions are important for the time series analysis of the solar wind and solar corona when $ U_0  $ and $ B_0 $ are comparable, e.g., for Parker Solar  Probe (PSP) when it is  close to the Sun. Hence, PSP's data provides an unique opportunity for testing the above predictions.    We hope to validate these predictions in near future. 

Solar wind data reveal another interesting property. Many authors\cite{Matthaeus:JGR1982rugged,Podesta:ApJ2007,Kasper:PRL2021} observe that the kinetic energy spectrum ($ E_u(k) $) is steeper than the magnetic energy spectrum ($ E_b(k) $). This relative steepening of $ E_u(k) $ with relative to $ E_b(k) $ is attributed to the energy transfers from the kinetic energy to the magnetic energy\cite{Verma:Fluid:2021,Verma:JPA2022}. Note that these energy transfers are critical for the magnetic field generation or dynamo. Interestingly, $ E^\pm(k) $ do not suffer from such steepening due to an absence of cross transfer between $ {\bf z}^+ $ and $ {\bf z}^-$ (see \citet{Verma:PR2004,Verma:book:ET}). This is another reason why $ E^\pm(k) $ are more reliable energy spectra compared to $ E_u(k) $ and $ E_b(k) $. It will be interesting to quantitatively compare the solar wind observations with the theoretical predictions made in this paper.

\acknowledgements
The author thanks  Rupak Mukherjee for useful suggestions and discussions. This work is partially supported by  the project CRG/2021/00l097 by  Science and Engineering Research Board (SERB), India.

%\bibliographystyle{abbrv_abhishek}
%\bibliographystyle{apsrev}
%\bibliography{/Users/mkv/Dropbox/docs-pub/bib/journal,/Users/mkv/Dropbox/docs-pub/bib/book,/Users/mkv/Dropbox/docs-pub/bib/book_chapter,/Users/mkv/Dropbox/docs-pub/bib/thesis} 

\end{document}